# Enhancing CdTe Solar Cell Performance by Reducing the "Ideal" Bandgap of CdTe through CdTe$_{1-x}$Se$_x$ Alloying

Jingxiu Yang [1,2] and Su-Huai Wei [2]*

[1] Department of Materials Science and Engineering, Jilin Jianzhu University, Changchun, 130118, China

[2] Beijing Computational Science Research Center, Beijing, 100193, China

Email: yangjingxiu@csrc.ac.cn; suhuaiwei@scrc.ac.cn

**Abstract**

CdTe is one of the leading materials for low cost, high efficiency thin-film solar cells, because it has a high absorption coefficient and a nearly ideal band gap of 1.48 eV for solar cell according to the Shockley-Queisser limit. However, its solar to electricity power conversion efficiency (PCE) is hindered by the relatively low open circuit voltage ($V_{OC}$) due to intrinsic defect related issues. Here, we propose the strategy of improving CdTe solar cell performance by *reducing* the "ideal" band gap of CdTe to gain more short-circuit current from long-wavelength absorption without sacrificing much $V_{OC}$. Alloying CdTe with CdSe seems to be the most appropriate approach to reduce the band gap because of the large optical bowing and relatively small lattice mismatch in this system, even though CdSe has larger band gap than CdTe. Using the first principle hybrid functional calculation, we find that the minimum band gap of the CdTe$_{1-x}$Se$_x$ alloy can be reduced from 1.48 eV at x=0 to 1.39 eV at x=0.32. We also show that the formation of the alloy can improve the defect property, for example, p-type doping of CdTe by Cu$_{Cd}$ can be greatly enhanced by the alloying effects.

# I. Introduction

CdTe is one of the leading material for low-cost, high-efficient, thin film solar cells due to its good optoelectronic property and the easy way to fabricate[1]. Although the power conversion efficiency (PCE) of the CdTe-based solar cell has so far reached to an impressive 22.1%, it is still much below the Shockley-Queisser limit (32%)[2]. The current PCE in the world-record solar cell is mainly limited by the small open-circuit voltage ($V_{OC}$), which is about 0.85 V compared to its band gap of 1.48 V at room temperature, as well as the relatively low short-circuit current ($J_{SC}$), which reaches about 28 mA/cm$^2$ compared to $J_{SC}$ = 30 mA/cm$^2$ under the Shockley-Queisser limit [1,3]. Currently, most efforts to improve CdTe-based solar cell efficiency have been trying to improve $V_{OC}$ instead of the $J_{SC}$ because of the large deficiency in $V_{OC}$. Some success has been achieved in increasing $V_{OC}$ by group V doping in CdTe [4]. However, it is still not clear whether such approach can obtain stable p-type absorbers because non-equilibrium doping process has to be used to improve the p-type doping [5]. On the other hand, one may increase the PCE by increasing $J_{SC}$, which can be easily achieved by reducing the band gap of CdTe to harvest more long-wavelength sunlight. For example, if the band gap is reduced from 1.48 eV to 1.35 eV, the ideal $J_{SC}$ is increased from 30 mA/cm$^2$ to ~36 mA/cm$^2$. Because $V_{OC}$ of the current champion CdTe solar cell is still much lower than the band gap[6,7], reducing the band gap of CdTe is not expected to cause much decrease of the $V_{OC}$.

Band gap tuning through alloying is widely used in semiconductors. Alloying CdTe at cation site could hardly achieve the reduction of the band gap, because the band gap always becomes wider when Cd is substituted by isovalent Zn[8,9], and it is not desired to try alloying HgTe with CdTe given the toxicity of Hg. Therefore, one can only try to reduce the band gap of CdTe through alloying CdTe at anion site. The band gap of CdS and CdSe is 2.52 eV and 1.74 eV, respectively[10]. Although the band gap of CdS and CdSe are both larger than that of CdTe, alloying CdS or CdSe into CdTe can effectively reduce its band gap due to the large bowing effect[11]. Because the lattice mismatch between CdS and CdTe is large, the solubility of S into CdTe is low, which has been confirmed by previous theoretical and experimental studies [11-

13]. Therefore, alloying CdTe with CdSe forming CdTe$_{1-x}$Se$_x$ seems to be the best choice to reduce the band gap effectively. Some of the recent experimental studies has already shown that diffusing CdSe into CdTe layer enables the increase of the J$_{SC}$ [2,14-16]. However, it is not clear how the band gap of CdTe$_{1-x}$Se$_x$ changes with the composition and what could be the minimum achievable band gap in this system to maximize the increase of the J$_{SC}$.

Furthermore, high p-type doping in CdTe is usually required for its solar cell performance, because as a minority carrier device, its electron mobility is much higher than the hole mobility. Although the dominant intrinsic p-type defect in CdTe is V$_{Cd}$, the obtained hole carrier density is too low for a good solar cell because V$_{Cd}$ has high formation energy. Therefore, extrinsic p-type dopants, such as Cu$_{Cd}$, is often used in commercial CdTe-based solar cells [17-19]. However, it is also not clear how the formation of CdTe$_{1-x}$Se$_x$ alloy affects the doping properties in CdTe.

In this work, using the first principle hybrid-functional calculations, we find that the minimum of the band gap of the CdSe$_x$Te$_{1-x}$ alloy can approach 1.39 eV at about x= 0.32. Our investigation of the doping property of the alloy reveals that the formation of the impurity Cu$_{Cd}$ exhibits dramatic bowing effect on the impurity formation energy, which can be utilized to improve the PCE. The obtained band structure and the defect properties of the CdSe$_x$Te$_{1-x}$ alloy suggest that CdSe$_x$Te$_{1-x}$ alloy should be a better solar cell absorber than CdTe for the thin film solar cell application.

## II. Computational Methods

The first principle calculation in this work is performed by the VASP code [20,21]. PAW psuedopotentials with an energy cutoff of 350 eV were employed. PBEsol functional[22] with GGA exchange correlation is used for the structure optimization of the bulk constitutes and alloys. All the atoms and the lattice vectors were fully relaxed until the force on each atom is less than 0.01 eV/Å. For the defect calculation, the lattice vectors of the optimized alloy is fixed with all the atoms inside the supercell relaxed. To calculate the band structures and the band offsets, we have employed the hybrid functional [23] consists of 32% exact Hartree-Fock exchange mixed with 68% PBE exchange with spin-orbit-coupling (SOC) to determine the band gap. This specific functional is chosen so that the calculated band gap of both zinc blende CdTe and CdSe are close to experimental values. Using the proposed functional, the calculated band gaps of zinc blende CdTe and CdSe are 1.52 eV and 1.69 eV, respectively, compared to the experiment values of 1.48 eV and 1.74 eV at room temperature [10]. The calculation of the band offsets of the series of $CdSe_xTe_{1-x}$ alloys follows the method described in our previous study [11].

The $CdSe_xTe_{1-x}$ alloy is assumed to be random and is mimicked by the special quasirandom structures (SQS)[24] in the cubic supercell of 512 or 64 atoms, when x=0, 0.25, 0.5, 0.75 and 1. The cubic supercell of 512 and 64 atoms are optimized with equivalent k-point sampling of $1\times1\times1$ and $2\times2\times2$, respectively. The averaged atomic correlation functions of the first neighbor pairs, triangles and tetrahedral of the SQS are the same as the perfect random alloys in the 512-atom supercells for all the mentioned concentrations. For the 64-atom supercell, the averaged atomic correlation functions of the first neighbored tetrahedral deviates from the perfect random alloys by 0.06 for x=0.25 and x=0.75, but is accurate enough for this case. The way to calculate the defect formation energy and the transition energy level is the same as stated in the previous work [25-27]. After testing with different functionals and supercells, the calculated formation energies are similar, and the calculated transition energy levels are converged to within 0.03 eV. Therefore, PBEsol functional and 64-atom supercells are adopted for the calculation of the doped alloys to reduce the computational cost.

### III. Results and Discussion

As described above, we have calculated the respective volume and mixing enthalpy $\Delta H_{mix}$ of the random $CdSe_xTe_{1-x}$ alloys with Se composition x=0, 0.25, 0.5, 0.75 and 1 as shown in Figure 1 (a) and (b). The obtained lattice constant is 6.55 and 6.13 Å for pure CdTe and CdSe, respectively, in reasonably good agreement with experiment data[10]. As x increases, the volume of the $CdSe_xTe_{1-x}$ alloy decrease linearly due to the smaller size of Se, following the Vegard's rule[28]. The $\Delta H_{mix}$ of the random alloy is defined as the energy difference between the CdSe$_x$Te$_{1-x}$ alloy and the pure CdSe and CdTe with the corresponding ratio. The calculated $\Delta H_{mix}$ can be described quite well by the quadratic function $\Omega x(1-x)$, with the interaction parameter $\Omega$ = 76.1 meV/f.u.. Using the calculated value of $\Omega$, the transition temperature is estimated to be 441K, which is much lower than the experimental growth temperature[16], therefore, it is easy to alloy CdSe into CdTe. In addition, the mixing enthalpy for x=0.25 is slightly lower than for x=0.75, reflecting the fact that it is easier to mix Se into CdTe than Te into CdSe.

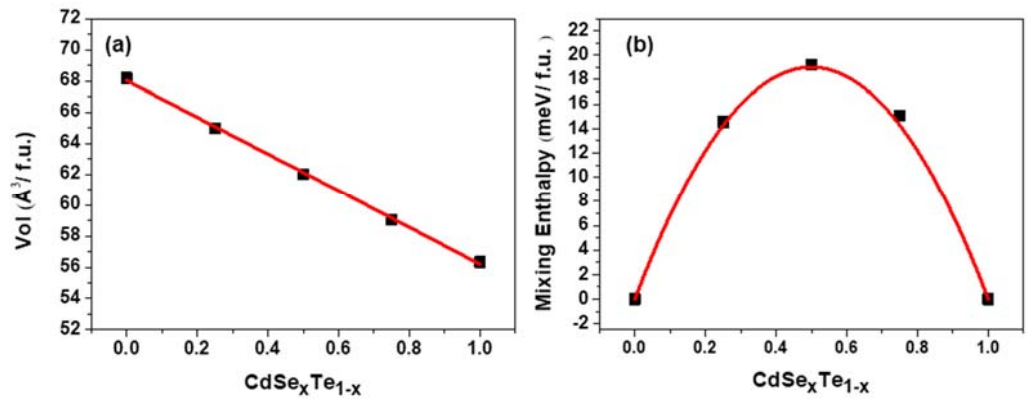

**FIG 1** The volume Å$^3$/f.u. (a) and the mixing enthalpy $\Delta H_{mix}$ meV/f.u. (b) as the function of the composition x for CdSe$_x$Te$_{1-x}$ alloys. The red lines in (a) and (b) are fitted curves.

The band gaps of the random CdSe$_x$Te$_{1-x}$ alloy are conventionally fitted to the equation:

$$E_g(CdSe_xTe_{1-x}) = (1-x)E_g(CdTe) + xE_g(CdSe) - bx(1-x) \qquad (1)$$

where b is the bowing coefficient for the band gap. The hybrid functional calculated band gaps as function of the composition x are plotted in Figure 2(a), where the band gap bowing parameter b is found to be 0.725 eV and the band gap minimum is found at x=0.38. Given the slight difference of the calculated and experimental band gaps, the composition for the band gap minimum also slightly varies. Using the calculated bowing parameter b=0.725 and the experimental value of the band gaps at room temperature, the obtained band gap minimum of the random CdSe$_x$Te$_{1-x}$ alloy is predicted to be 1.39 eV at x=0.32, in agreement with a recent experiment result[29].

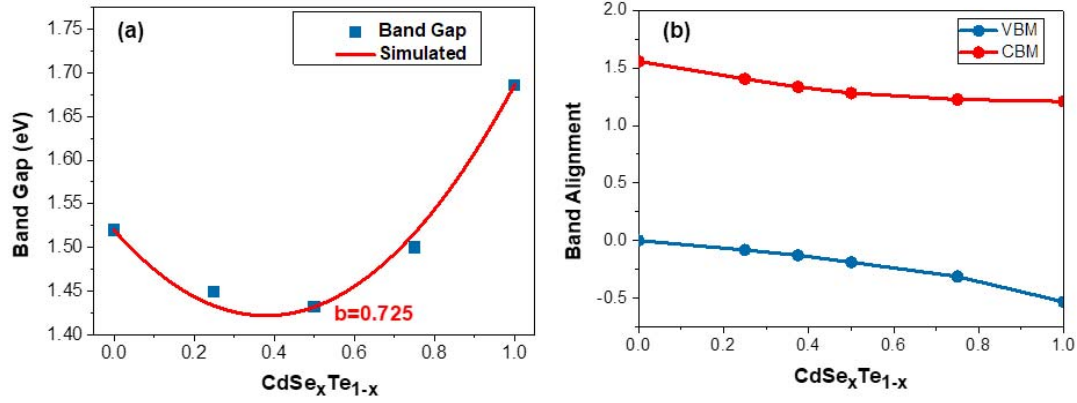

**FIG 2** (a) The calculated band gaps as function of x for CdSe$_x$Te$_{1-x}$ alloys; (b) The band alignments of the CdSe$_x$Te$_{1-x}$ alloys as function of x.

The bowing of the band gaps for CdSe$_x$Te$_{1-x}$ alloys is caused by the bowing of both the band edges. As shown in Figure 2 (b), the band offsets of the valence band minima (VBMs) and the conduction band minima (CBMs) between pure CdTe (x=0) and CdSe (x=1) are estimated to be 0.53 and 0.35 eV, respectively, consistent with the previous result [11]. Due to the strong intra valence band and intra conduction band coupling, the VBMs bow upwards and the CBMs bow downwards as x increases from 0 to 1, resulting the minimum band gap occurs at $x_{min}$. The type-II band alignment between CdTe and CdSe$_x$Te$_{1-x}$ suggests that a gradient CdSe$_x$Te$_{1-x}$ cell with Se-rich alloy in the front can help separate photogenerated electrons and holes, thus further improve the cell performance.

We first investigate the formation of the impurity Cu$_{Cd}$ in CdSe$_{0.375}$Te$_{0.625}$ alloy modeled by a 64-atom SQS containing all five type Se$_{4-n}$Te$_n$ (n=0-4) nearest neighbor motifs around each Cd atom. The formation energy of Cu$_{Cd}$ under Cd-rich condition at each possible site are calculated and plotted in Figure 3 (a). The formation energies of Cu$_{Cd}$ at charge state 0 and -1 depend mostly on the first neighbored configuration, although the farther neighbor configuration also has some effect, leading to the scattered formation energy within a given first neighbored motif. The averaged formation energies of the defect in different first neighbor motifs are shown in Figure 3 (b). It is obvious that the averaged formation energy increases as the number of Se atoms increase in its first neighbor. As more Se atoms surround the impurity in the first neighbor motif, the bonding orbitals of the impurity contains more Se 4p orbitals, which has lower orbital energy [Figure 2(b)], thus, to form $Cu_{Cd}^0$ state, it will cost more energy to create a hole. The formation energy of $Cu_{Cd}^{-1}$ (Figure 3(b) top) follows the trend of its neutral state (Figure 3(b) bottom), indicating the transition energy level $\varepsilon(0/-1)$ for $Cu_{Cd}$ is less sensitive to its local configuration compared to the neutral formation energy. In other word, the $Cu_{Cd}$ defect is more like a delocalized defect in $CdSe_xTe_{1-x}$ alloys.

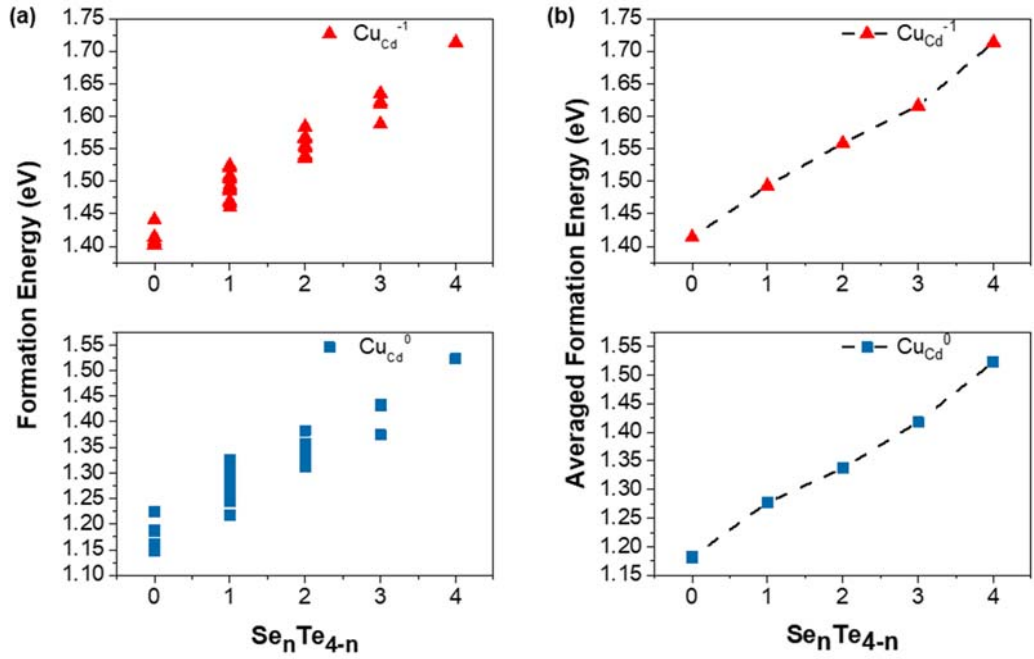

**FIG 3** (a) The formation energies of $Cu_{Cd}^{-1}$ and $Cu_{Cd}^0$ at each site in $CdSe_{0.375}Te_{0.625}$ alloy as a function of the number of the first neighbor Se atoms (**n**) around the impurity. (b) The arithmetic averaged formation energies of $Cu_{Cd}^{-1}$ and $Cu_{Cd}^0$ as a function of **n**. The black dashed line is just for guiding the eye. The Fermi level is set as 0 in both (a) and (b).

In alloys, the defect formation energy $\Delta H_f(\alpha, q, s, x)$ of defect $\alpha$ depends on charge state $q$, doping site $s$ and the alloy composition $x$. To statistically investigate the defect property, it is more convenient to introduce an effective formation energy[30] $\Delta H_{eff}(\alpha, q, x, T)$, which is $x$ and $T$ dependent weighted average of the formation energy as given in Eq. (2), where $k_B$ is the Boltzmann constant, and N is the total number of the corresponding defect sites in alloys. Obtaining the effective formation energy at charge states $0$ and $q$, we could also define the effective transition energy level $\varepsilon_{eff}(\alpha, 0/q, x, T)$ for defect $\alpha$, which is the Fermi energy at which defect $\alpha$ at charge state $0$ and $q$ has the same effective formation energy as shown in Eq (3).

$$exp[-\Delta H_{eff}(\alpha,q,x,T)/k_BT] = \frac{1}{N}\sum exp[-\Delta H_f(\alpha,q,s,x)/k_BT] \quad (2)$$

$$\varepsilon_{eff}(\alpha,0/q,x,T) = [\Delta H_{eff}(\alpha,0,x,T) - \Delta H_{eff}(\alpha,q,x,T) + qE_f]/q \quad (3)$$

Considering the limit condition for the effective formation energy, the Eq. (2) and Eq. (3) can be further deduced. At high temperature limit (T → ∞), all the sites has equal weight, thus the effective formation energy $\Delta H_{eff}$ ($\alpha$, $q$, $x$, ∞) is just the arithmetic average of the formation energies at all sites, so is the effective transition energy level $\varepsilon_{eff}$ ($\alpha$, $0/q$, $x$, ∞).

$$\Delta H_{eff}(\alpha,q,x,\infty) = \frac{1}{N}\sum \Delta H_f(\alpha,q,s,x);$$

$$\varepsilon_{eff}(\alpha,0/q,x,\infty) = \frac{1}{N}\sum \varepsilon_{eff}(\alpha,0/q,s,x). \quad (4)$$

On the other hand, at low temperature limit (T → 0), only the site with the lowest formation energy at charge $q$ ($s_0^q$) is occupied under equilibrium condition, so the effective formation energy $\Delta H_{eff}$ ($\alpha$, $q$, $x$, $0$) is just equal to $\Delta H_f$ ($\alpha$, $q$, $s_0^q$, $x$). The effective transition energy level $\varepsilon_{eff}$ ($\alpha$, $0/q$, $x$, $0$), therefore, is the energy difference between $\Delta H_f$ ($\alpha$, $0$, $s_0^0$, $x$) and $\Delta H_f$ ($\alpha$, $q$, $s_0^q$, $x$). Note that the $s_0^0$ and $s_0^q$ may not be at the same site.

$$\Delta H_{eff}(\alpha,q,x,0) = \Delta H_f(\alpha,q,s_0^q,x);$$

$$\varepsilon_{eff}(\alpha,0/q,x,0) = [\Delta H_f(\alpha,0,s_0^0,x) - \Delta H_f(\alpha,q,s_0^q,x) + qE_f]/q. \quad (5)$$

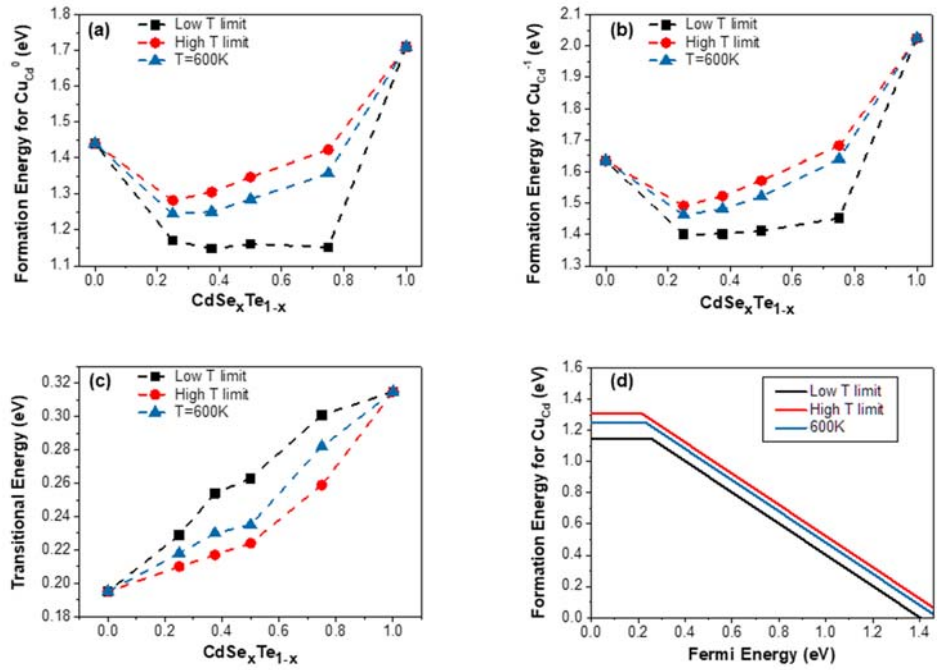

**FIG 4** The effective formation energies of $Cu_{Cd}^0$ (a), $Cu_{Cd}^{-1}$ (b) and the corresponding effective transition energy level (c) in $CdSe_xTe_{1-x}$ alloys (x=0, 0.25, 0.375, 0.5, 0.75 and 1) at the low temperature limit, the high temperature limit and a finite temperature T=600K. The Fermi level in (b) is set at 0. (d) The effective formation energy of $Cu_{Cd}$ as function of the Fermi energy in the $CdSe_{0.375}Te_{0.625}$ alloy at the low temperature limit, the high temperature limit and a finite temperature T=600K.

The calculated effective formation energies for the defect $Cu_{Cd}$ at neutral and -1 states in $CdSe_xTe_{1-x}$ alloys (x=0, 0.25, 0.375, 0.5, 0.75 and 1) at the low temperature limit, the high temperature limit and a finite temperature T=600K are shown in Figure 4 (a) and (b), respectively. It is interesting to see that the effective formation energy for $Cu_{Cd}$ impurity exhibits a large bowing, i.e., they are much smaller than that of the composition averaged values in the pure CdTe and the pure CdSe. This is because, in addition to the electronic effect discussed above, the strain effect also plays an important role. The formation of $Cu_{Cd}$ causes a compressive strain due to the smaller radius of Cu than Cd, thus the formation energy of $Cu_{Cd}$ will be reduced as the local volume surround Cu is reduced[31]. This is the case when Cu is surrounded by Te and $CuTe_4$ cluster is compressed in the $CdSe_xTe_{1-x}$ alloy, so the formation energy of $Cu_{Cd}^0$ is much lower in the $CdSe_xTe_{1-x}$ alloy than in pure CdTe. The formation energy of $Cu_{Cd}^0$ also decreases at the Se rich end when the $CuTe_4$ cluster is compressed most. At low temperature limit, Cu only occupy the lowest energy site ($CuTe_4$ cluster), so the bowing is the largest at the Se-rich side. At high temperature limit, the substitution occurs equally at all sites, so the effective formation energy change more smoothly as Se concentration increases. The formation of the $Cu_{Cd}^{-1}$ generally follows the trend of $Cu_{Cd}^{0}$ except that the bowing for $Cu_{Cd}^{-1}$ is less dramatic than the bowing for $Cu_{Cd}^{0}$ due to the larger size of the $Cu_{Cd}^{-1}$ impurity.

As expected, the effective transition energy level increases as Se concentration increases in the alloy. It is interesting to see in Figure 4 (c) that at a given composition the effective transition energy level decreases as the temperature increase. This is because at the low temperature, the site with lower formation energy is preferentially occupied, where the impurity energy level for $Cu_{Cd}^{0}$ is usually high to easily creating the hole. Therefore, the transition energy level (0/-1) is relatively high. At the high temperature limit, all the defect sites have nearly equal occupation probability, so the averaged effective transition energy is reduced. However, the *variation* of the effective transition energy is small at a given composition (~0.04 eV), reflecting that $Cu_{Cd}$ is a relatively delocalized defect in $CdSe_xTe_{1-x}$ alloys.

The formation energy of the $Cu_{Cd}^{0}$ defect in the $CdSe_{0.375}Te_{0.625}$ alloy range from 1.31 eV to 1.15 eV at Cd-rich limit with the transition energy level varying from 0.217 eV to 0.254 eV, depending on the synthetic temperature, as shown in Figure 4 (d). The insensitivity of the transition energy level and the lower formation energy of $Cu_{Cd}$ in the $CdSe_{0.375}Te_{0.625}$ alloy suggests Cu doping in the alloy is more effective than that in pure CdTe.

## IV. Conclusion

In summary, using first-principles calculations, we show that alloying CdTe with CdSe to form $CdSe_xTe_{1-x}$ alloys could be an effective approach to increase the PCE of the CdTe based thin film solar cells. The $CdSe_xTe_{1-x}$ alloy has two merits compared to CdTe: (1) reduced band gap (estimated to be 1.39 eV at x=0.32) to improve long-wavelength light harvest, thus improving $J_{SC}$ without significant effect on achievable $V_{OC}$; (2) lower formation energy of the shallow defect $Cu_{Cd}$ to improve the p-type conductivity. Experimental tests of our predictions are called for.

## Acknowledgements

This work was supported by the National Key Research and Development Program of China under Grant No. 2016YFB0700700 and the National Nature Science Foundation of China under Grant No. 51672023; 11634003; U1530401. We also acknowledge the computational support from the Beijing Computational Science Research Center.

[1]  M. A. Green, Y. Hishikawa, E. D. Dunlop, D. H. Levi, J. Hohl-Ebinger, and A. W. Y. Ho-Baillie, Solar cell efficiency tables (version 51), Prog. Photovolt. Res. Appl. **26**, 3 (2018).

[2]  A. R. Duggal, J. J. Shiang, W. H. Huber, and A. F. Halverson, (Google Patents, 2014).

[3]  R. M. Geisthardt, M. Topic, and J. R. Sites, Status and potential of CdTe solar-cell efficiency, IEEE J. Photovolt. **5**, 1217 (2015).

[4]  J. M. Burst *et al.*, CdTe solar cells with open-circuit voltage breaking the 1 V barrier, Nat. Energy **1**, 16015 (2016).

[5]  J.-H. Yang, W.-J. Yin, J.-S. Park, J. Burst, W. K. Metzger, T. Gessert, T. Barnes, and S.-H. Wei, Enhanced p-type dopability of P and As in CdTe using non-equilibrium thermal processing, J. Appl. Phys. **118**, 025102 (2015).

[6]  G. M. A., E. Keith, H. Yoshihiro, W. Wilhelm, and D. E. D., Solar cell efficiency tables (version 40), Prog. Photovolt. Res. Appl. **20**, 606 (2012).

[7]  G. M. A., E. Keith, H. Yoshihiro, W. Wilhelm, and D. E. D., Solar cell efficiency tables (version 46), Prog. Photovolt. Res. Appl. **23**, 805 (2015).

[8]  S.-H. Wei and A. Zunger, Disorder effects on the density of states of the II-VI semiconductor alloys $Hg_{0.5}Cd_{0.5}Te$, $Cd_{0.5}Zn_{0.5}Te$, and $Hg_{0.5}Zn_{0.5}Te$, Phys. Rev. B **43**, 1662 (1991).

[9]  Y. Wang, J. Xu, P. Ren, X. Zhuang, H. Zhou, Q. Zhang, X. Zhu, and A. Pan, Complete composition tunability of $Cd_{1-x}Zn_xTe$ alloy nanostructures along a single substrate, Mater. Lett. **105**, 90 (2013).

[10]  *Semiconductors: Basic Data* (Spinger, Berlin, 1982), 2nd ed.

[11]  S.-H. Wei, S. B. Zhang, and A. Zunger, First-principles calculation of band offsets, optical bowings, and defects in CdS, CdSe, CdTe, and their alloys, J. Appl. Phys. **87**, 1304 (2000).

[12]  X. Yang *et al.*, Preparation and characterization of pulsed laser deposited CdS/CdSe bi-layer films for CdTe solar cell application, Mater. Sci. Semicond. Process. **48**, 27 (2016).

[13]  B. E. McCandless, G. M. Hanket, D. G. Jensen, and R. W. Birkmire, Phase behavior in the CdTe–CdS pseudobinary system, J. Vac. Sci. Technol., A **20**, 1462 (2002).

[14]  N. R. Paudel and Y. Yan, Enhancing the photo-currents of CdTe thin-film solar cells in both short and long wavelength regions, Appl. Phys. Lett. **105**, 183510 (2014).

[15]  N. R. Paudel, J. D. Poplawsky, K. L. Moore, and Y. Yan, Current Enhancement of CdTe-Based Solar Cells, IEEE J. Photovolt. **5**, 1492 (2015).

[16]  J. D. Poplawsky, W. Guo, N. Paudel, A. Ng, K. More, D. Leonard, and Y. Yan, Structural and compositional dependence of the $CdTe_xSe_{1-x}$ alloy layer photoactivity in CdTe-based solar cells, Nat. Commun. **7**, 12537 (2016).

[17]  J. Perrenoud, L. Kranz, C. Gretener, F. Pianezzi, S. Nishiwaki, S. Buecheler, and A. N. Tiwari, A comprehensive picture of Cu doping in CdTe solar cells, J. Appl. Phys. **114**, 174505 (2013).

[18]  L. Kranz *et al.*, Doping of polycrystalline CdTe for high-efficiency solar cells on flexible metal foil, Nat. Commun. **4**, 2306 (2013).

[19]  D. L. Bätzner, A. Romeo, H. Zogg, R. Wendt, and A. N. Tiwari, Development of efficient and stable back contacts on CdTe/CdS solar cells, Thin Solid Films **387**, 151 (2001).

[20]  G. Kresse and J. Furthmüller, Efficiency of ab-initio total energy calculations for metals and semiconductors using a plane-wave basis set, Comput. Mater. Sci. **6**, 15 (1996).

[21]  G. Kresse and J. Furthmüller, Efficient iterative schemes for ab initio total-energy calculations using a plane-wave basis set, Phys. Rev. B **54**, 11169 (1996).

[22]  J. P. Perdew, A. Ruzsinszky, G. I. Csonka, O. A. Vydrov, G. E. Scuseria, L. A. Constantin, X. Zhou, and K. Burke, Restoring the density-gradient expansion for exchange in solids and surfaces, Phys. Rev.

Lett. **100**, 136406 (2008).

[23] A. V. Krukau, O. A. Vydrov, A. F. Izmaylov, and G. E. Scuseria, Influence of the exchange screening parameter on the performance of screened hybrid functionals, J. Chem. Phys. **125**, 224106 (2006).

[24] S.-H. Wei, L. G. Ferreira, J. E. Bernard, and A. Zunger, Electronic properties of random alloys: Special quasirandom structures, Phys. Rev. B **42**, 9622 (1990).

[25] J. Ma, S.-H. Wei, T. A. Gessert, and K. K. Chin, Carrier density and compensation in semiconductors with multiple dopants and multiple transition energy levels: Case of Cu impurities in CdTe, Phys. Rev. B **83**, 245207 (2011).

[26] J.-H. Yang, S. Chen, H. Xiang, X. G. Gong, and S.-H. Wei, First-principles study of defect properties of zinc blende MgTe, Phys. Rev. B **83**, 235208 (2011).

[27] J.-H. Yang, W.-J. Yin, J.-S. Park, J. Ma, and S.-H. Wei, Review on first-principles study of defect properties of CdTe as a solar cell absorber, Semicond. Sci. Technol. **31**, 083002 (2016).

[28] L. Vegard, Die Konstitution der Mischkristalle und die Raumfüllung der Atome, Zeitschrift für Physik **5**, 17 (1921).

[29] M. Lingg, A. Spescha, S. G. Haass, R. Carron, S. Buecheler, and A. N. Tiwari, Structural and electronic properties of $CdTe_{1-x}Se_x$ films and their application in solar cells, Sci. Technol. Adv. Mater. **19**, 683 (2018).

[30] J. Ma and S.-H. Wei, Bowing of the defect formation energy in semiconductor alloys, Phys. Rev. B **87**, 241201 (2013).

[31] J. Zhu, F. Liu, G. B. Stringfellow, and S.-H. Wei, Strain-Enhanced Doping in Semiconductors: Effects of Dopant Size and Charge State, Phys. Rev. Lett. **105**, 195503 (2010).